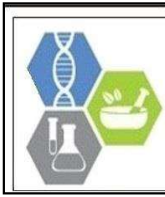

# Evaluation of *in vitro* antibacterial activity and phytochemical profile of aqueous leaf extract of *Asystasia variabilis*

## R Wijerathna, NAV Asanthi, WD Ratnasooriya, RN Pathirana and NRM Nelumdeniya

**R Wijerathna**
Department of Pharmacy, Faculty of Allied Health Sciences, General Sir John Kotelawala Defence University, Werahera, Sri Lanka

**NAV Asanthi**
Department of Pharmacy, Faculty of Allied Health Sciences, General Sir John Kotelawala Defence University, Werahera, Sri Lanka

**WD Ratnasooriya**
Department of Basic Sciences, Faculty of Allied Health Sciences, General Sir John Kotelawala Defence University, Werahera, Sri Lanka

**RN Pathirana**
Department of Basic Sciences, Faculty of Allied Health Sciences, General Sir John Kotelawala Defence University, Werahera, Sri Lanka

**NRM Nelumdeniya**
Department of Pharmacy, Faculty of Allied Health Sciences, General Sir John Kotelawala Defence University, Werahera, Sri Lanka

**Correspondence**
R Wijerathna
Department of Pharmacy, Faculty of Allied Health Sciences, General Sir John Kotelawala Defence University, Werahera, Sri Lanka

**Abstract**
This study evaluated the *in vitro* antibacterial effect and the phytochemical profile of aqueous extract of fresh mature leaves of *Asystasia variabilis*, a Sri Lankan indigenous plant, against four common wound infective bacteria (*Staphylococcus aureus*, *Bacillus subtilis*, *Pseudomonas aeruginosa* and *Escherichia coli*) using Kirby-Bauer disk diffusion test. Gentamicin 10 μg/ disk and distilled water was used as positive and negative controls respectively The study revealed, for the first time, that the extract possessed significant antibacterial activity against all four test organisms in a concentration dependent manner (r values ranging from 0.921-0.992, $P<0.01$) with inhibition zone diameters ranging between 8 and 28 mm. Highest antibacterial activity was exhibited against *B. subtilis* at 1000 μg/ disk (27.43±0.02 mm). The extract showed inhibitory effects comparable to gentamicin towards *B. subtilis* and *P. aeruginosa* at 500 μg/ disk and towards *E. coli* and *S. aureus* at 1000 μg/ disk. Qualitative phytochemical screening revealed the presence of flavonoids, tannins, phenols, cardiac glycosides, amino acids, carbohydrates, alkaloids and saponins. Therefore it is likely that the antibacterial effect of the extract is mediated by synergistic mechanisms. Furthermore, results of this study scientifically justified the claim traditional and folk medicine in the treatment of abscesses, wounds and ulcers and indicated the potential for the development of a novel drug from mature leaves of *Asystasia variabilis*.

**Keywords:** *Asystasia variabilis*, antibacterial activity, wound infective bacteria, phytochemical profile

## 1. Introduction
Infectious diseases are caused by microorganisms such as bacteria, viruses, parasites or fungi and can be spread, directly or indirectly, from one person to another [1]. Infections are rampant in developing countries due to factors such as unsanitary living conditions and malnutrition [2]. Therefore, the use of antimicrobial agents has a drastic impact on improving the health of the general population. An antimicrobial drug is a substance that inhibits the growth or destroy microbes in the human body without causing damage to the host cells [3]. These substances may be produced synthetically or can be extracted from microbes or natural sources such as plants. However, although numerous synthetic antibiotics have been discovered and are currently being used, they induce side effects [4] and the irresponsible use of these antibiotics have given rise to antibiotic resistance [5]. Therefore this issue has become one of the main concerns in modern medicine. Furthermore, discoveries of novel antibiotic classes have not been made since 1987 resulting in a "discovery void" [6]. Therefore, because of these reasons, there is a crucial need to develop new antibacterial agents. In Sri Lankan traditional and folk medicine a large number of plants are claimed to possess antibacterial activity and therefore used to treat wounds, abscesses and ulcers. However, although Sri Lankan traditional medicine makes use of numerous indigenous herbs to treat diseases, nowadays many important medicinal plants are considered as weeds due to unawareness of the people. One such medicinal plant is *Asystasia variabilis* which belongs to the family Acanthaceae. It is known as "Gada Puruk" in Sinhala, and has a distribution in Sri Lanka and South India. It commonly grows under shade by footpaths and streams in secondary forests of the moist lowlands and mid country. The flowering season of *Asystasia variabilis* is from August to March. The stem of the plant is erect at first, then semi-scandent, sharply grooved on opposite sides and slightly tumid above nodes. Leaves can be variable in shape from oval-elliptic to broadly oval-lanceolate, with size ranging from 3.5cm-14.5cm × 0.6cm-3.4cm. Further leaves possess an acute base tapering into petiole which is acuminate and glabrous. Corolla comprises of a pale pinkish violet tube, bent at the base and faintly purple within the throat. Dotted red, the lobes are ovate-oblong and unequal, where the mid lobe is broader with two raised ridges going down tube. Shape of the seeds vary from ovate-deltoid to.

orbicular and are 3-4.5mm broad and shallowly tuberculate [7]. A. variabilis is used to treat ailments such as abscesses, wounds and ulcers in Sri Lankan traditional and folk medicine [8], although validity of this usage has not yet been scientifically justified. Although previous studies had not been carried out to evaluate the antibacterial activity of A. variabilis, prior studies carried out with regards to other species belonging to the genus Asystasia has revealed the presence of antimicrobial activity and phytochemicals such as flavonoids and saponins [9, 10, 11]. Hence, this present study was aimed at investigating the in vitro antibacterial activity of aqueous extract of mature leaves of *Asystasia variabilis* on common wound infective bacteria (Escherichia coli, Pseudomonas aeruginosa Staphylococcus aureus, and Bacillus subtilis) and a qualitative phytochemical analysis was carried out to determine its phytoconstituents.

## 2. Materials and Methods
### 2.1 Collection and authentication
Whole plant specimens of *Asystasia variabilis* were collected from Bulathsinhala, situated in Colombo District, Western Province, Sri Lanka (GIS cordinates-6.6487° N, 80.1791° E) in August, 2017. The plant material was identified and authenticated by a botanist at the National Herbarium, Peradeniya, Sri Lanka.

### 2.2 Preparation of aqueous leaf extract of *Asystasia variabilis*
Fresh mature leaves (100g) were first washed with distilled water and cut into small pieces and macerated in 100 ml of distilled water using mortar and pestle. The resultant mixture was kept at 4 oC for 24 hours in an airtight, sterile container and was filtered firstly by a double muslin cloth, secondly by Whatman No.1 filter paper, and thirdly by a membrane filter (0.48 μm pore size) to obtain the final extract.

### 2.3 Determination of concentration of original extract
Three 25 ml portions of the final extract was subjected to steam evapouration in previously weighed petri dishes (w1) and the weight of each petri dish with the residue was recorded (w2). The concentration was calculated by dividing the weight of the residue (w2-w1) by volume of the sample (25 ml). The average of the three values was taken as concentration of the final extract.

### 2.4 In vitro antibacterial assay
Antibacterial activity of aqueous extract of fresh mature leaves of A. variabilis was evaluated using Kirby-Bauer disk diffusion method [12] against E. coli (ATCC 25922), B. subtilis (ATCC 6633), P. aeruginosa (ATCC 27853) and S. aureus (ATCC 25923). Suspensions of the test bacterial strains were adjusted to 0.5 McFarland turbidity standard ($1.5 \times 10^8$ CFU) and inoculated by streaking on Muller Hinton Agar. The original extract was diluted with distilled water to prepare a concentration series of 12.5, 25.0, 37.5 and 50.0 mg/mL. Four sterilized filter paper disks (Whatman 4) were impregnated with 20 μL of the different leaf extract concentrations such that the disks contain 250 μg/ disk, 500 μg/ disk, 750 μg/ disk, 1000 μg/ disk. Distilled water was used as the negative control while gentamicin 10 μg/ disk was used as the positive control. The impregnated discs were aseptically placed on the agar surface at equidistance. The plates were allowed to stand for 20 minutes at room temperature (25 oC) to pre diffuse and incubated at 37 oC for 24 hours and the inhibition zone diameters was measured using a vernier caliper to nearest 0.02 mm. The experiment was carried out in triplicates.

### 2.5 Phytochemical screening
The aqueous extract of *Asystasia variabilis* was subjected to following qualitative phytochemical tests. Analysis were performed for alkaloids using Dragendorff's, Mayer's, Wagner's and Hager's tests and for anthraquinones using Modified Borntrager's test. Test for carbohydrates was achieved via Molisch's, Benedict's and Fehling's tests and for cardiac glycosides via Keller Killani test. Foam test was carried out for saponins, while Liebermann Burchard test was carried out for steroids. Moreover Salkowski test was performed for terpenoids while Shinoda's and Sodium Hydroxide test were performed for flavonoids. Gelatin test was performed for tannins, while Ferric Chloride and Lead Acetate tests were performed for detection of phenols. Biuret test and Ninhydrin test were performed for the investigation of proteins and amino acids respectively [13, 14].

### 2.6 Statistical analysis of results
Data were expressed as mean ± S.E.M. and linear regression analysis was used to determine dose dependencies using SPSS version 23 software. Significance was set at $P<0.01$.

## 3. Results
In vitro antibacterial activity of aqueous leaf extract of A. variabilis and positive control against test organisms are depicted in Table 1. As shown, the highest susceptibility was shown against B. subtilis with maximum inhibition zone of 27.43±0.02 mm at 1000 μg/ mL followed by P. aeruginosa. Lowest activity was evident with E. coli and S. aureus. The extract showed antibacterial activity comparable to gentamicin towards B. subtilis and P. aeruginosa at 500 μg/ disk (22.63±0.02 mm and 15.91±0.02 mm respectively) and towards E. coli and S. aureus at 1000 μg/ disk (18.96±0.01 mm and 17.69±0.04 mm respectively). The results of linear regression (Table 2) revealed that the antibacterial activities were dose dependent for all four test organisms. Qualitative phytochemical analysis of A. variabilis revealed the presence of tannins, phenols, cardiac glycosides, alkaloids, carbohydrates, flavonoids, saponins, and amino acids and the absence of steroids, proteins, anthraquinones and terpenoids (Table 3). All qualitative tests for alkaloids indicated a positive result except for Mayer's test.

**Table 1:** In vitro antibacterial screening results of A. variabilis aqueous leaf extract (Mean ± S.E.M.)

| Test organism | Zones of inhibition (mm) | | | | |
|---|---|---|---|---|---|
| | 250 μg/disk | 500 μg/disk | 750 μg/disk | 1000 μg/disk | Gentamicin 10 μg/disk |
| E. coli | 10.40±0.12 | 12.33±0.18 | 15.67±0.13 | 18.96±0.01 | 18.60±0.12 |
| S. aureus | 10.79±0.02 | 15.57±0.04 | 15.98±0.03 | 17.69±0.04 | 17.17±0.10 |
| P. aeruginosa | 8.49±0.01 | 15.91±0.02 | 18.01±0.02 | 21.36±0.03 | 14.66±0.02 |
| B. subtilis | 14.61±0.02 | 22.63±0.02 | 26.02±0.01 | 27.43±0.02 | 19.63±0.02 |

**Table 2:** Evaluation of the correlation between concentration and antibacterial activity of *A. variabilis*

| Microorganism | r value (P < 0.01) |
|---|---|
| *E. coli* | 0.992 |
| *P. aeruginosa* | 0.964 |
| *B. subtilis* | 0.941 |
| *S. aureus* | 0.921 |

**Table 3:** Phytochemical analysis of A. variabilis aqueous leaf extract

| Secondary etabolites | Aqueous leaf extract |
|---|---|
| Alkaloids | + |
| Carbohydrates | + |
| Anthraquinones | - |
| Cardiac Glycosides | + |
| Saponins | + |
| Steroids | - |
| Terpenoids | - |
| Phenolic compounds | + |
| Tannins | + |
| Flavonoids | + |
| Proteins | - |
| Amino acids | + |

### 4. Discussion

The present study examined, for the first time, the in vitro antibacterial activity of fresh mature leaves of *Asystasia variabilis* against two Gram- positive (B. subtilis and S. aureus) and two Gram- negative (E. coli and P. aeruginosa) bacteria commonly associated with wound infection, using standard Kirby-Bauer disk diffusion method, a bio assay which is a widely used, simple, reliable, inexpensive, validated and sensitive technique in evaluating antibacterial activity of herbal extracts [15] and synthetic drugs [16].

To our knowledge, yet the antibacterial activity and qualitative phytochemical profile of A. variabilis has not been evaluated. An aqueous fresh leaf extract was used to assess the antibacterial activity since an extract prepared by macerating fresh mature A. variabilis leaves in water is generally used in traditional and folk medicine as a topical application for wounds, abscesses and ulcers. Furthermore, it is generally recommended to use water in testing the pharmacologic activity of phytomedicines, since most traditional medicine is prepared in a similar manner, although methanolic extracts are extensively used in the evaluation of in vitro antibacterial activity of herbal extracts [9, 10]. The difference in potencies between extracts prepared using solvents of varying degree of polarity could be due to the variation in solubility of the bioactive compounds in the particular solvent.

Under experimental conditions, the results clearly showed marked antibacterial activity comparable to the positive control gentamicin 10 μg/ disk against all test organisms. The results also revealed to be dose dependent, indicating that the inhibitory effect is genuine, causal and specific. B. subtilis was indicated to be the most susceptible bacterial species followed by P. aeruginosa, E. coli and S. aureus. However, mild antibacterial activity was exhibited in prior studies involving other species belonging to the genus Asystasia [9, 10, 11]. The discrepancy observed in the degree of inhibition could be due to the difference in species and environmental factors such as pH of soil, humidity, temperature, irrigation and rainfall which changes the quality and quantity of phytochemicals in the plant, thereby changing the bioactivity [17]. In addition, variations in methods of extraction, bio assay and utilization of different solvents may have contributed to the difference in inhibitory effect.

In the current study, the phytochemical analysis of aqueous leaf extract of *Asystasia variabilis* revealed the presence of flavonoids, tannins, phenols, cardiac glycosides, amino acids, carbohydrates, alkaloids and saponins and the absence of steroids, anthraquinones, proteins and terpenoids. Mayer's test did not show a positive result although all the other qualitative tests for alkaloids (Hager's test, Wagner's test, Dragendorff's test) exhibited a positive response even when the test was repeated several times at high concentrations. To our knowledge, all results obtained from this entire research study are novel findings for the plant *Asystasia variabilis*. According to previous studies, the antibacterial activity of secondary metabolites such as flavonoids and tannins are attributed to mechanisms such as formation of complexes with the bacterial cell wall, causing damage to the cell membrane and enzyme inactivation whereas phenols cause cell membrane disruption. Defensive action against bacteria is provided by alkaloids via intercalation with the microbial cell wall and DNA [18] while saponins act by permeabilization of the microbial cell membrane [19]. Hence, it is likely that the antibacterial effect of the extract may be a synergistic effect resulted by multiple mechanisms of action of several secondary metabolites [20], a characteristic which lowers the incidence of developing antimicrobial resistance. Findings of this research study showed that leaves of A. variabilis possessed strong antibacterial activity against B. subtilis, E. coli, S. aureus, P. aeruginosa which confirms the great potential of secondary metabolites. To our knowledge, no previous scientific studies have been conducted to evaluate the antimicrobial effect of A. variabilis and all results of the present investigation are novel findings for A. variabilis. Outcomes of this study scientifically justify the claim made by Sri Lankan traditional and folk medicine that mature leaves of A. variabilis is an effective treatment for wounds, abscesses and ulcers and thereby indicate the potential of developing a new antibiotic with fewer side effects from an indigenous plant that is readily available and grows irrespective of specific climatic conditions.

### 5. Conclusion

In conclusion, this study revealed for the first time, that the aqueous extract of mature leaves of A. variabilis possessed in vitro antibacterial activity against common wound infective bacteria (E. coli, B subtilis, P. aeruginosa and S. aureus). Highest inhibitory effect was exhibited against B subtilis. Qualitative phytochemical analysis revealed the presence of several secondary metabolites previously proven to possess antimicrobial effects. It is likely the antibacterial action of the extract is a synergistic effect of these bioactive compounds. All results of this investigation are novel findings. In addition, the results justify the use of *Asystasia variabilis* in the treatment of abscesses, wounds and ulcers in traditional and folk medicine and indicate the potential of developing new antibacterial agents with low incidence of resistance from a plant most people are unaware of and mistakenly destroyed as a weed.

### 6. Acknowledgements
We would like to thank Dr. Ranjith. J. Welikala for the guidance given and the assistance of Mrs. Renuka Rajapaksha is greatly appreciated.

7. References
1. Bannister B, Gillespie S, Jones J. Infection: Microbiology and Management. Edn 3, John Wiley & Sons, London, 2009, 3-5.
2. Zaidi AKM, Awasthi S, Janaka deSilva H. Burden of infectious diseases in South Asia. BMJ: British Medical Journal. 2004; 328(7443):811-815.
3. Vasanthakumari R. Textbook of Microbiology. BI Publications Pvt Ltd, New Delhi, 2007, 80-82.
4. Cunha BA. Antibiotic side effects. Medical Clinics. 2001; 85(1):149-85.
5. http://www.who.int/en/news-room/fact-sheets/detail/antibiotic-resistance. 15 Jan, 2018.
6. Silver LL. Challenges of antibacterial discovery. Clinical Microbiology Reviews. 2011; 24(1):71-109.
7. Dassanayake MD. A Revised Handbook of the Flora of Ceylon. Routledge, 2017; 7:79.
8. http://www.instituteofayurveda.org/plants/plants_detail.php?i=5&s=Scientific_name, 2018.
9. Kumar S, Kumar V, Shashidhara S, Chandrashekhar MS. Antioxidant and antimicrobial activities of Asystasia dalzelliana: A novel plant. Journal of Pharmacy Research. 2011; 4(1):186-188.
10. Hamid AA, Aiyelaagbe OO, Ahmed RN, Usman LA, Adebayo SA. Preliminary phytochemistry, antibacterial and antifungal properties of extracts of Asystasia gangetica Linn T. Anderson grown in Nigeria. Adv Appl Sci Res. 2011; 2(3):219-226.
11. Hamid AA, Aiyelaagbe OO. Pharmacological investigation of Asystasia calyciana for its antibacterial and antifungal properties. Int. J Chem. Biochem. Sci. 2012; 1:99-104.
12. Bauer AW, Kirby WM, Sherris JC, Turck M. Antibiotic susceptibility testing by a standardized single disk method. American Journal of Clinical Pathology. 1966; 45(4):493.
13. Evans W, Trease G. Trease and Evans Pharmacognosy. Edn 16, Saunders, Edinburg, 2004.
14. Roopashree TS, Dang R, Rani RS, Narendra C. Antibacterial activity of antipsoriatic herbs: Cassia tora, Momordica charantia and Calendula officinalis. International Journal of Applied Research in Natural Products. 2008; 1(3):20-28.
15. Barbour E, Al Sharif M, Sagherian V, Habre A, Talhouk R, Talhouk S et al. Screening of selected indigenous plants of Lebanon for antimicrobial activity. Journal of Ethnopharmacology. 2004; 93(1):1-7.
16. Jones R, Ballow C, Biedenbach D. Multi-laboratory assessment of the linezolid spectrum of activity using the Kirby-Bauer disk diffusion method: Report of the Zyvox Antimicrobial Potency Study (ZAPS) in the United States. Diagnostic Microbiology and Infectious Disease. 2001; 40(1-2):59-66.
17. Bernillon S, Biais B, Deborde C, Maucourt M, Cabasson C, Gibon Y et al. Metabolomic and elemental profiling of melon fruit quality as affected by genotype and environment. Metabolomics. 2013; 9(1):57-77.
18. Cowan MM. Plant products as antimicrobial agents. Clinical Microbiology Reviews. 1999; 12(4):564-582.
19. Moses T, Papadopoulou and Osbourn A. Metabolic and functional diversity of saponins, biosynthetic intermediates and semi-synthetic derivatives. Critical Reviews in Biochemistry and Molecular Biology. 2014; 49(6):439-462.
20. Alabi OA, Haruna MTH, Anokwuru CP, Jegede T, Abia H, Okegbe VU, Esan BE et al. Comparative studies on antimicrobial properties of extracts of fresh and dried leaves of Carica papaya (L) on clinical bacterial and fungal isolates. Advances in Applied Science Research. 2012; 3(5):3107-3114.